\documentclass[twocolumn,prl,aps,superscriptaddress,showpacs]{revtex4-1}

\usepackage{dcolumn}

\usepackage{mathptmx}  
\usepackage{bm}         

\usepackage[english]{babel}

\usepackage[pdftex]{graphicx} 
\usepackage{float}
\usepackage{epstopdf}
\usepackage{color}
\definecolor{mygrey}{gray}{0.45}
\definecolor{myblue}{rgb}{0.2,0.2,0.8}
\definecolor{myzard}{cmyk}{0,0,0.05,0}
\definecolor{mywhite}{rgb}{1,1,1}
\definecolor{myred}{rgb}{1,0.,0.3}
\usepackage[colorlinks=true,citecolor=myblue,linkcolor=myred,hidelinks=true]{hyperref}

\usepackage{latexsym}
\usepackage{amsmath,amssymb,amsfonts,amsbsy,mathtools}
\usepackage{upgreek}    
\usepackage{mathrsfs}   
\usepackage{cancel}     
\usepackage{mathdots}   
\usepackage{setspace}   
\usepackage{dsfont}     

\usepackage{comment}
\usepackage{enumerate}  
\usepackage{appendix}
\usepackage{booktabs}
\usepackage{natbib}

\begin{document}

\title{A deep learning approach to resonant light transmission through single subwavelength apertures} 
\author{David Alonso-González}
\affiliation{Departamento de F\'{i}sica Te\'{o}rica de la Materia Condensada and Condensed Matter Physics Center (IFIMAC), Universidad Aut\'{o}noma de Madrid, E-28049 Madrid, Spain}
\author{Michel Frising}
\affiliation{Departamento de F\'{i}sica de la Materia Condensada and Condensed Matter Physics Center (IFIMAC), Universidad Aut\'{o}noma de Madrid, E-28049 Madrid, Spain}
\author{Ferry Prins}
\affiliation{Departamento de F\'{i}sica de la Materia Condensada and Condensed Matter Physics Center (IFIMAC), Universidad Aut\'{o}noma de Madrid, E-28049 Madrid, Spain}
\author{Jorge Bravo-Abad}
\affiliation{Departamento de F\'{i}sica Te\'{o}rica de la Materia Condensada and Condensed Matter Physics Center (IFIMAC), Universidad Aut\'{o}noma de Madrid, E-28049 Madrid, Spain}

\begin{abstract}
Resonant transmission of light is a surface-wave assisted phenomenon that enables funneling light through subwavelength apertures milled in otherwise opaque metallic screens. In this work, we introduce a deep learning approach to efficiently compute and design the optical response of a single subwavelength slit perforated in a metallic screen and surrounded by periodic arrangements of indentations. First, we show that a semi-analytical framework based on a coupled-mode theory formalism is a robust and efficient method to generate the large training datasets required in the proposed approach. Second, we discuss how simple, densely connected artificial neural networks can accurately learn the mapping from the geometrical parameters defining the topology of the system to its corresponding transmission spectrum. Finally, we report on a deep learning tandem architecture able to perform inverse design tasks for the considered class of systems. We expect this work to stimulate further work on the application of deep learning to the analysis of light-matter interaction in nanostructured metallic films.

\end{abstract}

\maketitle

The advance of the technological abilities for structuring metals in the micro- and nanometer length scales led in 1998 to the experimental discovery that the optical transmission through an array of subwavelength holes milled in a metallic screen can, for certain wavelengths, be significantly boosted with respect to what was predicted for independent holes~\cite{eot1}. Later, it was shown that a single aperture can also feature enhanced transmission, along with remarkable beaming effects, when surrounded by periodic arrangements of surface corrugations~\cite{eot2}. The underlying physical origin of both phenomena relies on the excitation of electromagnetic resonances (surface plasmons) running along the corrugated metallic surfaces~\cite{eot3,eot4}. Importantly, the transmission can be precisely engineered by varying the geometry of the surface corrugations, allowing for spectrally, directionally, and polarization selective transmission properties~\cite{laux2008,jun2011,davis2017,deleo2017}. 

On a different front, during the last decades \textit{deep learning} (DL), a sub-area of machine learning~\cite{goodfellow}, has drawn a great deal of attention from the scientific community due to its ability to deal with complex data-driven problems~\cite{cirac}. DL techniques are based on \emph{training} \textit{artificial neural networks} (ANNs), a collection of nonlinear units or nodes called \textit{artificial neurons}~\cite{murphy}. Each artificial neuron receives a \emph{signal}, processes it, and then transmits a new output to other neurons connected to it. The signal goes from an input layer formed by a set of neurons to an output one, usually after traversing several layers of neurons (so-called \emph{hidden layers}).

In this work, we report on the application of ANNs to the simulation of the optical response of a single subwavelength aperture surrounded by periodic arrangements of indentations and perforated in an opaque metallic screen. In our approach, we first use a semi-analytical theoretical framework, based on a well-known coupled-mode theory, to generate large enough data sets than can be used to train the studied ANNs. Then, we show that ANNs can indeed be used to accurately obtain the non-linear mapping between the topology of the studied class of nanophotonic structures and their corresponding transmission spectra. In addition, we discuss a deep-learning based inverse design algorithm that enables finding the geometrical parameters of the system that yield a desired optical response.

We focus on the structure sketched in Fig.~1, which have been extensively analyzed in the context of resonant transmission phenomena~\cite{eot4}. It is based on a single slit (of width $b$), perforated on a metallic film (of thickness $W$) flanked to the left and right by finite arrays of one-dimensional rectangular indentations. Each of these arrays is characterized by its corresponding periodicity ($d_1$ and $d_2$ label the periodicity of the left and right arrays, respectively), and the width and depth of each indentation. In what follows, we assume that all indentations have the same width ($a$) but different depths in the left and right arrays ($h_1$ and $h_2$, respectively). We also assume the whole system is embedded in air (refractive index $n=1$).  The number and specific degrees of freedom chosen here to describe the geometry allows having a moderate number of free parameters, while at the same time capturing with their variation most of the phenomenology present in this class of devices.

\begin{figure}[b]
    \centering
    \includegraphics[width=8.5cm]{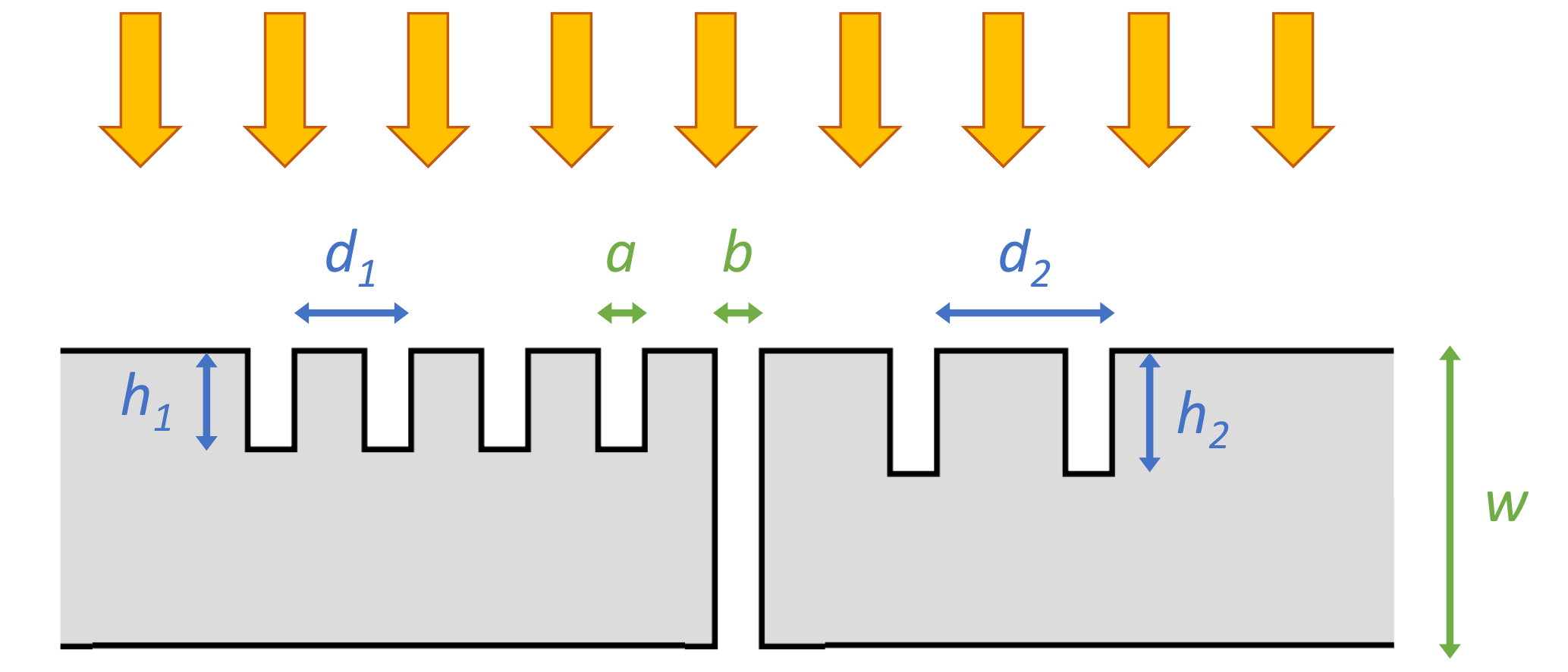}
    \caption{Schematics of the system under study, along with the geometrical parameters defining it. Yellow arrows on top represent the plane wave illumination incident normally onto the system.}
    \label{fig:structure}
\end{figure}

One of the key ingredients of any deep learning approach is the dataset used to train the algorithmic architecture. For studying our specific problem, we have generated the dataset using a coupled mode theory (CMT) framework. This theoretical framework has been widely applied to the studied systems and it is well-known for its accuracy and computational efficiency \cite{eot4}. The latter aspect is crucial for our purposes, as will be key for having access to large training sets required in the proposed approach.

We consider the case of $p$-polarized incident light, for which the most apparent resonant transmission effects have been reported. Within a CMT framework, electromagnetic (EM) fields are expanded by a set of plane waves in air and by a modal expansion inside the slit and indentations. As we are mainly interested in subwavelength apertures, we consider that only the fundamental propagating eigenmode is excited. We assume that the metal behaves as a perfect metal, expelling the electric field from its bulk. This assumption, ideally suited for applying a CMT as the one used here, captures in a semi-quantitative fashion most enhanced transmission phenomena in the optical regime~\cite{eot3}. Within this formalism, the total transmittance of the studied setup can be computed through the following compact expression: $ T= \textrm{Im}\left\{ E^{I} E^{O*} \right\} / \sin (kW)$, where $E^{I}$ and $E^{O}$ are complex amplitudes related the $x$-component of the electric field at illuminated and non-illuminated the exit of the slit, respectively, and $k=2\pi/\lambda$, with $\lambda$ being the wavelength in vacuum of the incident light (see Ref.~\cite{eot4}, and references therein, for details on these magnitudes and a description of the re-illumination effects between the indentations and the slit).

\begin{figure}[t]
    \centering
    \includegraphics[width=8.3cm]{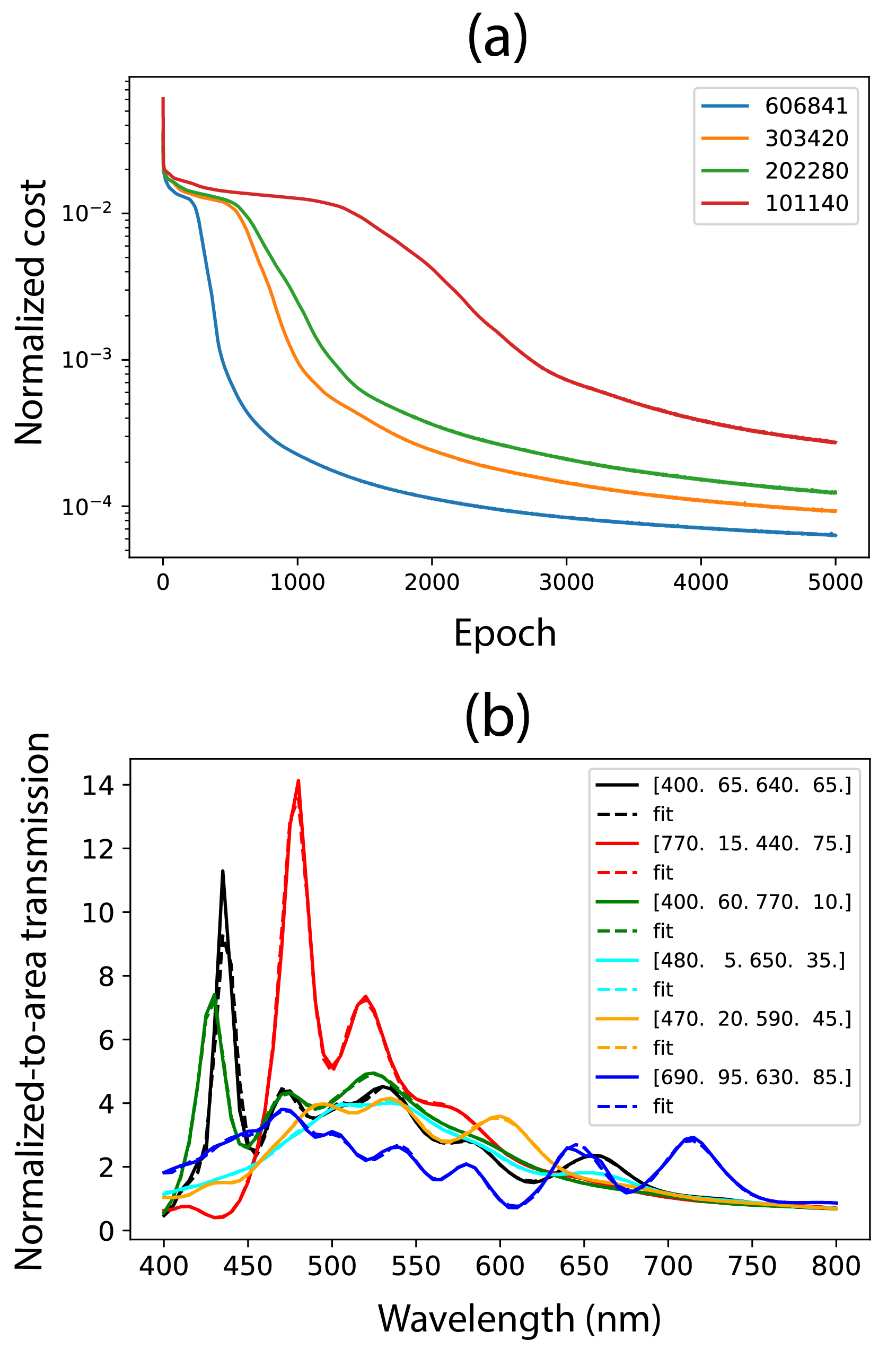}
    \caption{(a) Evolution of the cost function with epochs in the optimal ANN, as obtained for different sizes of the dataset, ranging from 606,481 (blue line) to 101,140 (red line). (b) Predictions of the ANN for a random set of six transmission spectra that were contained in the test set (i.e., not used to train the network). The values of the parameters corresponding to each case (in nm) are indicated in the legend following the notation [$d_1$ $h_1$ $d_2$ $h_2$]. Solid lines and dashed correspond, respectively, to the results computed with a coupled-mode theory formalism and the predictions from the ANN.}
    \label{fig:train4param_size}
\end{figure}

First, we trained our neural network to accurately predict the transmission spectra of the studied class of systems. In order to generate the corresponding dataset, we varied the periodicity ($d_1$ and $d_2$) and the depth ($h_1$ and $h_2$) of both left and right sides of the illuminated part of the structure. In all cases, we fixed the number of indentations to the left and right of the slit to $N=10$ (i.e., we have a total of 20 indentations surrounding the slit). Following Ref.~\cite{soljacic}, the initial architecture of the network consisted of 3 hidden layers with 250 neurons each. The input layer has 4 neurons corresponding to the four input parameters ($d_1$, $d_2$, $h_1$ and $h_2$), while the output layer has 81 neurons, corresponding to a sampling of the transmission spectra at 81 different wavelength values, equally spaced between $400$~nm and $800$~nm. Our dataset is therefore composed by sets of the four parameters (playing the role of \emph{features} in our deep learning approach) and their corresponding transmission spectra (these spectra being the corresponding \emph{labels}). 

The detailed count of samples forming this dataset is obtained as follows. We vary independently the periodicities of the left and right arrays of indentations from $400$~nm to $800$~nm with a step of $10$~nm, yielding 41 possible values for the periodicity of each array. We also vary independently the depths of the indentations of the left and right arrays, from $5$~nm to $95$~nm with a step of $5$~nm, producing 19 possible configurations for each side of the slit. Therefore, we have a total of $(41\times19)^2=606,841$ spectra in the complete dataset. From this dataset, we take a 80\% as the training set, and a 10\% each for the validation and the test sets.

Next, we performed a systematic set of calculations to find the optimal hyperparameters of the described ANN. In all these studies, to quantify the difference between the ANN's predictions and the CMT results, during the training process we used a cost function based on the mean average error (MAE) of all sampling points assumed in the spectra. Regarding the architecture of the network, we found that our initial guess (including 3 hidden layers and 250 neurons per layer) leads to an optimal performance of the network. On the other hand, from all the activation functions, a \emph{Sigmoid} was the one yielding the best accuracy (see~\cite{SI} for details on this analysis). The best performing optimizer among the series we tried resulted to be \emph{Adam}. For that optimizer we tried several learning rates, $\eta$. We checked that values of $\eta$ lower than $10^{-4}$ show a very good performance, and found a particularly good performance for $\eta=3 \times 10^{-5}$. In this study of hyperparameters, the variation in the efficiency of the ANN's training with the size of the training set is of particular interest. Figure~2(a) summarizes the numerical results for the evolution of the cost function with the number of epochs, for different sizes of the training set. As seen, as the size of the training set decreases, the training process becomes slower and for the maximum number of epochs considered in this work (5,000 epochs) the cost function increases significantly when considering training sets smaller than approximately 200,000 samples. This result shows that the ability to access large enough datasets is a key ingredient to enable the proposed approach.

Figure 2(b) shows the predictions of the ANN featuring the optimal hyperparameter set and for the largest number of epochs and training set sizes considered in this work (5,000 and 606,841, respectively). A set of six randomly selected transmission spectra from the test set is considered for illustration. As observed, for the displayed set, the predictions of the ANN (solid lines) show very good agreement with the results obtained with the CMT theory (dashed lines), to the extent that in most cases the results from both approaches are indistinguishable to the naked eye. A more quantitative way to assess the performance of the network can be obtained through the evaluation of the average value of the cost function over the main data subsets involved in this problem (namely, the training, validation and test sets). Table I summarizes the corresponding results, showing that an average error of 0.42\% is obtained for the three datasets (note that the fact that the validation and test set features the same value of the cost function also shows that no overfitting is occurring in this case).

\begin{table}[b]
  \centering
    \begin{tabular}{|c|c|c|c|c|c|}
    \toprule
                &            & \multicolumn{3}{c|}{Average of the cost function (\%)} \\
    \midrule
    Epochs     & Samples    & Train      & Validation & Test \\
    \midrule
    5000       & 606841     & 0.42       & 0.42       & 0.42 \\
    \bottomrule
    \end{tabular}%
  \caption{Values of the cost function computed for the main data subsets of the analyzed deep learning approach.}
  \label{tab:finalcosts}%
\end{table}%


\begin{figure}[t]
    \centering
    \includegraphics[width=8.3cm]{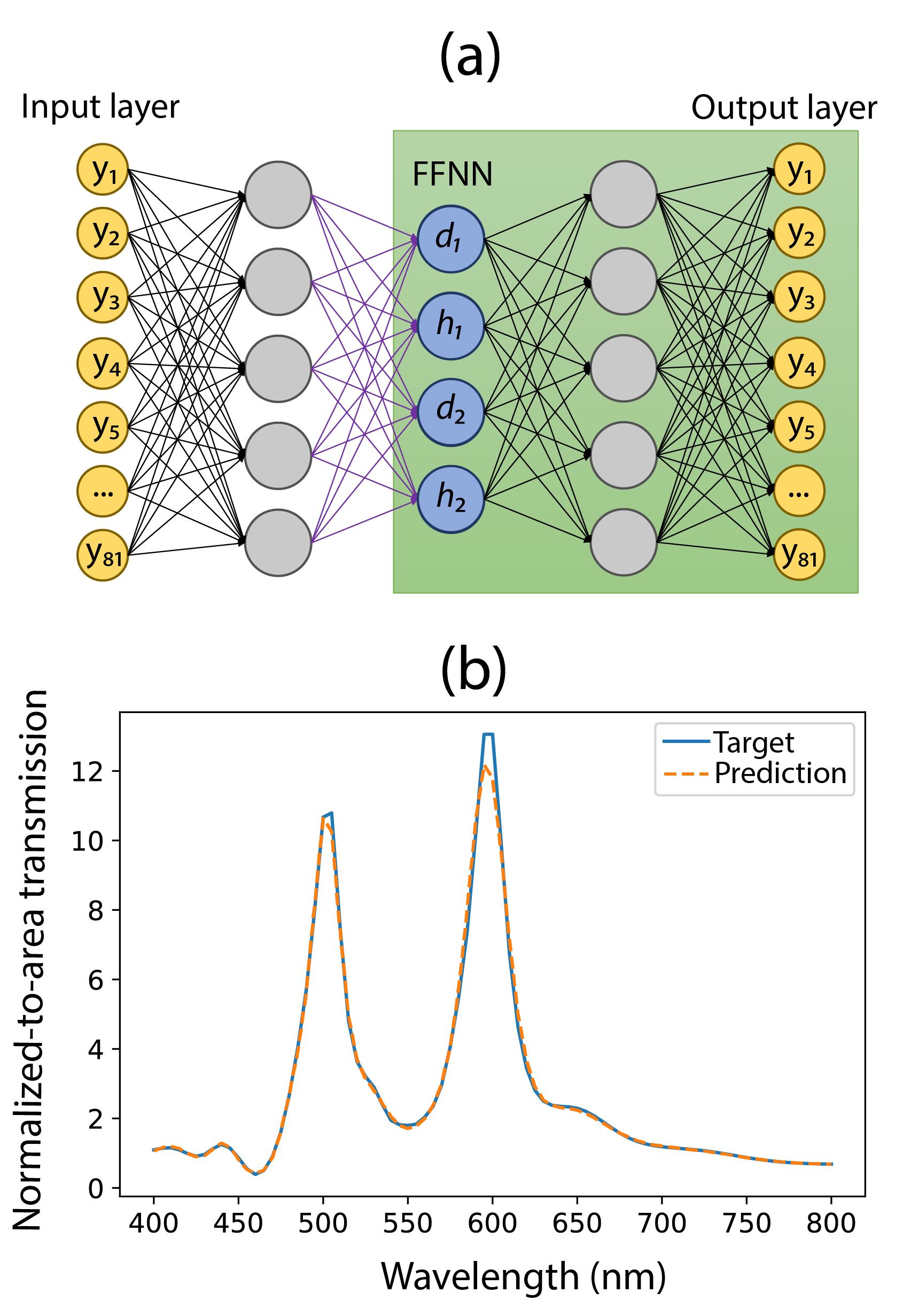}
    \caption{(a) Schematics of the tandem artificial neural network used for inverse design. Yellow circles to the left and right represent the input and output layers, respectively (both holding a sampling of 81 points of the spectra). Gray circles represent the hidden layers of the proposed ANN architecture. Blue circles correspond to the inner artificial neurons holding values that define the geometry of the structure. The area in green marks the part of the deep learning system that corresponds to a pretrained feed-forward neural network (FFNN). Trainable connections among neurons are displayed in purple. (b) Comparison between a targeted spectrum and the prediction of the ANN for than spectrum, obtained by using the deep learning architecture displayed in (a).}
\end{figure}

Now, we turn to the \emph{inverse design} of nanophotonic devices~\cite{so,fan}. The main idea underlying deep learning for inverse design can be explained in simple terms: it consists of ANNs able to design nanophotonic systems that can match a predefined \emph{targeted} optical response. In this work, inspired by Refs.~\cite{so,fan} (and references therein), we consider the tandem architecture shown in Fig. 3(a), which we have specifically tailored to perform inverse design tasks for the studied kind of metallic structures. The artificial neurons of the input layer (shown in yellow in Fig.~3(a)) hold the sampling of the targeted transmission spectrum (as in the previous ANN we assume a 81-point sampling of the spectrum). The output layer has a similar structure, but in this case the neurons hold the sampling of the spectrum produced by the whole deep learning system. The overall loss function of this deep learning system is set up to minimize the difference between the targeted spectrum and the one obtained from the output layer (see details in~\cite{SI}). Right at the middle of the system we have inserted a small hidden layer, consisting in only 4 neurons (shown in blue in Fig.~3(a)). Importantly, those neurons hold the values of the geometrical parameters characterizing the structure (the values of the set $[d_1 ~ h_1 ~ d_2 ~ h_2]$) that, when operating in inference mode, the proposed deep learning system predicts as optimal for achieving the targeted response. The connection of this inner layer with the input and output layers (to the left and right, respectively, in the sketch of Fig.~3(a)) is done through the stacking of hidden layers (represented as columns of grey circles). We assume the same topology for these hidden layers as the one used in Fig.~2, i.e., 3 hidden layers of 250 artificial neurons are set to the left and right of the inner layer.

To demonstrate the performance of the tandem architecture shown in Fig.~3(a), we have selected as target spectrum the one shown in Fig.~3(b) (blue line). It correspond to a physical setting, not included in the training set, defined by the geometrical parameters: $[d_1 ~ h_1 ~ d_2 ~ h_2]=[470~75~560~95]$~nm. As observed, the spectrum features two resonant transmission peaks at very different wavelengths that are close in magnitude and width. We believe that reaching a solution for a spectrum of these characteristics, which combines simultaneously many different aspects of the physical mechanisms underlying resonant transmission in the considered structures~\cite{eot4}, is a demanding proof-of-principle task that illustrates well the inverse design capabilities of the system sketched in Fig.~3(a). 

Remarkably, we have found that to deploy the above described tandem architecture as efficient inverse design engine, it is not necessary to train the whole deep learning system shown in Fig.~3(a). An alternative approach consists in \emph{transferring} to the layers placed to the right of the central inner layer of the tandem structure (green area in Fig.~3(a)) the weights of the optimized predictive ANN discussed in the first part of this work (labeled as \emph{feed-forward neural network, FFNN} in Fig.~3(a)). In addition, we transfer to the layers located to the left of the central inner layer the same weights, but with an important difference: we do not transfer the ones corresponding to the connections between last hidden layer of that part of the architecture and the inner central layer (these connections are shown in purple in Fig.~3(a)). Following this route it is possible to avoid the large computational load associated to a detailed training of the whole system (see~\cite{SI} for the full details of the implementation of this numerical process).

By running the well-known backpropagation algorithm~\cite{goodfellow} in the above discussed setup, we obtain that the system shown in Fig.~3(a) is able to predict that the targeted spectrum corresponds to the set of geometrical parameters $[469.96 ~~ 75.25 ~~ 559.96 ~~ 95.18]$~nm, which are very close to the ones associated to the targeted spectrum ($[470 ~ 75 ~ 560 ~ 95]$~nm). Figure~3(b) shows the spectrum, also produced by the same tandem structure, for the predicted optimal set of parameters (orange dashed line). The very good agreement between predicted and targeted spectra demonstrates the high degree of accuracy that can be obtained with the proposed inverse design approach.

To sum up, in this work we have studied the application of deep learning techniques to the simulation and design of the optical response of a single subwavelength aperture milled in an opaque metallic screen and surrounded by periodic arrangements of indentations. In particular, we have employed a densely connected ANN to accurately obtain the non-linear mapping between the topology of this class of structures and their corresponding transmission spectra. To do that, we have used a semi-analytical coupled mode theory that has allowed us to access the large dataset needed in the training process of our ANN. We have found very good agreement between the transmission spectra predicted by the proposed ANN and the spectra obtained with the coupled-mode theory. We have also explored an important application of these kind of methods, namely, the inverse design of the system. Specifically, we have proposed a tailored ANN tandem architecture as solution for inverse design problems in this context. We believe the present study can contribute to a wider adoption of deep learning techniques as efficient tools for the analysis of light-matter interaction in metallic films structured at the subwavelength scale.


\begin{references}
 \bibitem{eot1} \textsc{T.W. Ebbesen, H.J. Lezec, H.F. Ghaemi, T. Thio, P.A. Wolff}, \textit{Extraordinary optical transmission through subwavelength hole arrays}, Nature 391, 667 (1998).
    \bibitem{eot2} \textsc{H.J. Lezec, A. Degiron, E. Devaux, R.A. Linke, L. Martín-Moreno, F.J. García-Vidal and T.W. Ebbesen}, \textit{Beaming light from a Subwavelength Aperture}, Science, 297, 820 (2002).
    \bibitem{eot3} \textsc{L. Martín-Moreno, F.J. García-Vidal, H.J. Lezec, K.M. Pellerin, T. Thio, J.B. Pendry and T.W. Ebbesen}, \textit{Theory of extraordinary optical transmission through subwavelength hole arrays}, Phys. Rev. Lett. 86, 1114 (2001).
    \bibitem{eot4} \textsc{F.J. García-Vidal, L. Martín-Moreno, T.W. Ebbesen and L. Kuipers}, \textit{Light passing through subwavelength apertures}, Rev. Mod. Phys. 82, 729 (2010).
    \bibitem{laux2008} \textsc{E. Laux, C. Genet, T. Skauli, T.W. Ebbesen}, \textit{Plasmonic photon sorters for spectral and polarimetric imaging}, Nat. Phot. 2, 161 (2008).
    \bibitem{jun2011} \textsc{Y.C. Jun, K.C.Y. Huang, M.L. Brongersma}, \textit{Plasmonic beaming and active control over fluorescent emission}, Nat. Comm. 2, 283 (2011).
    \bibitem{davis2017} \textsc{M.S. Davis, W. Zhu, T. Xu, J.K. Lee, H.J. Lezec, A. Agrawal}, \textit{Aperiodic nanoplasmonic devices for directional colour filtering and sensing}, Nat. Comm. 8, 1347 (2017) .
    \bibitem{deleo2017} \textsc{E. De Leo, A. Cocina, P. Tiwari, L.V. Poulikakos, P. Marqués-Gallego, B. le Feber, D.J. Norris, F. Prins}, \textit{Polarization Multiplexing of Fluorescent Emission Using Multiresonant Plasmonic Antennas}, ACS Nano 11, 12167 (2017).
    \bibitem{goodfellow} \textsc{I. Goodfellow, Y. Bengio, A. Courville}, \textit{Deep Learning}, The MIT Press (Cambridge, 2016).
    \bibitem{cirac} \textsc{G. Carleo, I. Cirac, K. Cranmer, L. Daudet, M. Schuld, N. Tishby, L. Vogt-Maranto, and L. Zdeborova}, \textit{Machine learning and the physical sciences}, Rev. Mod. Phys. 91, 045002 (2019). 
    \bibitem{murphy} \textsc{K.P. Murphy}, \textit{Machine Learning: A Probabilistic Perspective}, The MIT Press (Cambridge, 2012).
    \bibitem{soljacic} \textsc{J. Peurifoy, Y. Shen, L. Jing, Y. Yang, F. Cano-Renteria, B.G. DeLacy, J.D. Joannopoulos, M. Tegmark, M. Soljačić}, \textit{Nanophotonic particle simulation and inverse design using artificial neural networks}, Science Advances 4, eaar4206 (2018).
    \bibitem{so} \textsc{S. So, T. Badloe, J. Noh, J. Bravo-Abad, J. Rho}, \textit{Deep learning enabled inverse design in nanophotonics}, Nanophotonics 9, 1041 (2020).
    \bibitem{fan} \textsc{J. Jiang, M. Chen, J.A. Fan}, \textit{Deep neural networks for the evaluation and design of photonic devices}, Nat. Rev. Mat. 1, 22 (2020). 
    \bibitem{SI} \textrm{See detailed description in the Supplement accompanying this work}.
\end{references}
\end{document}